\documentclass[twocolumn,showpacs,amsfonts]{revtex4-1}
\usepackage{amsmath}
\usepackage{latexsym}
\usepackage{float}
 \usepackage{amssymb}
\usepackage{graphicx}
\usepackage{textcomp}
\usepackage{hyperref}
\usepackage{subfigure}

\def\ba{\begin{eqnarray}}
\def\ea{\end{eqnarray}}
\def\be{\begin{equation}}
\def\ee{\end{equation}}
\def\bm{\begin{math}}
\def\me{\end{math}}

\newcommand{\dummy}

\begin{document}
\title{Dimension Dependence of Clustering Dynamics in Models of Ballistic Aggregation and Freely Cooling Granular Gas}
\author{ Subhajit Paul and Subir K. Das$^{*}$}
\affiliation{Theoretical Sciences Unit, Jawaharlal Nehru Centre for Advanced Scientific Research,
 Jakkur P.O., Bangalore 560064, India}

\date{\today}

\begin{abstract}
~~Via event-driven molecular dynamics simulations we study kinetics of clustering in assemblies of inelastic particles 
in various space dimensions. We consider two models, viz., the ballistic aggregation model (BAM) and the freely cooling granular gas 
model (GGM), for each of which we quantify the time dependence of kinetic energy and average mass of clusters (that form 
due to inelastic collisions). These quantities, for both the models, exhibit power-law behavior, at least in the long time limit. For the 
BAM, corresponding exponents exhibit strong dimension dependence and follow a hyperscaling relation. 
In addition, in the high packing fraction limit the behavior of 
these quantities become consistent with a scaling theory that predicts an inverse relation between energy and mass. 
On the other hand, in the case of the GGM we do not find any evidence for 
such a picture. In this case, even though the energy decay, irrespective of packing fraction, 
matches quantitatively with that for the high packing fraction picture of the BAM, it is inversely proportional to 
the growth of mass only in one dimension, and the growth appears to be rather insensitive to the choice of the dimension, unlike the BAM.
\end{abstract}


\maketitle

\section{ Introduction}
~~Growth in many physical situations occurs due to inelastic collisions among aggregates 
\cite{carne_5, naim1_5, naim2_5, trizac1_5, frache_5, pias_5, mid1_5, bril3_5, binder1, binder2, roy1, shimi, gold_5, rogers_5, nie_5, matt_5, takada_5, rai_5, rangoli_5}. 
Typical examples \cite{carne_5, trizac1_5, bril1_5, binder1, binder2, roy1, shimi, rogers_5, mid1_5, matt_5, takada_5, rai_5, spahn_5, bril2_5} 
are growth of liquid droplets and solid clusters in upper atmosphere, 
clustering in cosmic dust, etc. In this context, a simple model, referred to as the ballistic aggregation model (BAM) \cite{carne_5, trizac1_5, trizac2_5}, 
has been 
of much theoretical interest. In this model, spherical particles move with constant velocities and merge upon collisions to form larger aggregates,
 by keeping the shape unchanged. In this process, mass and momentum of the system remain conserved, whereas the (kinetic) energy decays.  
It is, of course, understood that 
following collisions fractal structures will emerge \cite{mid1_5, paul1_5} in space dimension $d>1$. 
Even though appears a bit unrealistic from that point of view, this simple model can provide important insights into the understanding of growth 
in many complex systems \cite{carne_5, trizac1_5, frache_5, pias_5, trizac2_5}.
In fact, in many situations colliding objects undergo deformation 
and so, if the collision interval is long, the above mentioned spherical structural approximation is reasonably good.
\par
~~There has been longstanding interest in understanding of decay of energy ($E$) and growth of average mass ($m$) in the BAM.
Carnevale, Pomeau and Young (CPY) \cite{carne_5}, via scaling arguments, predicted that
\begin{equation}\label{ba_scaling}
 m\sim \frac{1}{E} \sim t^{2d/d+2}.
\end{equation}
An inherent assumption in arriving at this quantitative picture is that the particle (or cluster) momenta are uncorrelated \cite{pathak2_5}. 
Even though the predictions in Eq. (\ref{ba_scaling}) are in agreement with the computer simulations in $d=1$, discrepancies 
have been reported \cite{trizac1_5, paul1_5} for $d>1$. Another theory in this context, by Trizac and Hansen \cite{trizac1_5},
 predicts the existence of a hyperscaling relation involving the time dependence of energy and mass. If one writes
\begin{equation}\label{energ_exp}
 E \sim t^{-\theta}
\end{equation}
and 
\begin{equation}\label{mass_exp}
 m \sim t^{\zeta},
\end{equation}
then the (positive) power-law exponents $\theta$ and $\zeta$ are expected to be connected to each other in $d$ dimensions via \cite{trizac1_5}
\begin{equation}\label{hyper_scl}
 2\zeta + d\theta =2d.
\end{equation}
\par
~~While Eq. (\ref{ba_scaling}) satisfies the hyperscaling relation in Eq. (\ref{hyper_scl}), the former prediction is expected to be true, 
as stated above, 
when cluster momenta are uncorrelated, i.e., when collision frequency is high \cite{trizac2_5, paul1_5}. 
This latter picture will be valid when the particle density is reasonably large. 
\par
~~There have been efforts to confirm Eq. (\ref{hyper_scl}). Such works \cite{trizac1_5, paul1_5}, 
however, restricted attention to $d=2$. In this work we undertake a comprehensive 
study, by considering a wide range of density and adopting an accurate method of analysis, to check the validty of Eq. (\ref{hyper_scl}) in $d=2$ and $3$. In this 
process we also intend to quantify the convergence with respect to the validity of Eq. (\ref{ba_scaling}), in the above mentioned dimensions. 
\par
~~Furthermore, there have been works \cite{paul1_5, pathak1_5, naim3_5} that aim to understand if the freely cooling granular gas model (GGM) is equivalent to the BAM.
 Here note that in the case of GGM \cite{gold_5}, the coefficient of 
restitution ($e$) lies in the range $0<e<1$. Thus, in this model, following inelastic collisions, particles do not merge to form bigger particles. 
Nevertheless, velocity parallelization 
occurs due to reduction in normal relative velocity, following the collisions. This gives rise to clustering phenomena in this model
\cite{gold_5, nie_5, paul1_5, pathak1_5, shinde1_5, naim3_5, pathak2_5, das1_5, das2_5, paul2_5}. In majority of the  previous 
studies \cite{nie_5, pathak1_5, naim3_5} with GGM, 
the primary objective was to quantify the decay of energy and it has been shown \cite{carne_5, paul1_5, pathak1_5} that the decay is consistent 
with Eq. (\ref{ba_scaling}), a prediction for the BAM, in all dimensions. 
There also exist reports on the growth of mass and other aspects \cite{paul1_5, das1_5, paul2_5}. In $d=1$ equivalence, 
with respect to decay of energy, growth of mass and related aging, between the two models has been established \cite{paul1_5, 
naim3_5, shinde1_5}. It has also been pointed out \cite{paul1_5, pathak1_5, paul2_5} that such a picture may not exist in higher dimensions. 
An appropriate understanding of dimension dependence of growth of mass, however, is still lacking. 
In this work, we undertake a study with that objective.
\par
~~The objectives of this paper, thus, are to investigate the correctness of the hyperscaling relation of 
Eq. (\ref{hyper_scl}), for the BAM in $d=2$ and $3$,  
and check if at least an analogous picture exists for the connection between the decay of energy and the growth of mass in the GGM. 
To verify the hyperscaling relation \cite{trizac1_5} in the BAM, 
we consider different packing fractions. 
We observe that the relation is valid irrespective of the dimension and packing fraction. In the high density limit, in addition, 
the prediction of Eq. (\ref{ba_scaling}) appears correct. This is because of the fact, as already mentioned,  
that the collisions are more random and thus velocities 
are uncorrelated in high density situation. On the other hand, the results for the GGM does not provide any hint of the existence 
of a relation of this type.
\par
~~The rest of the paper is organised as follows. In Section II we provide more details on the theoretical predictions for the BAM. Models and 
methods are discussed in Section III. Results are presented in Section IV. Finally, Section V concludes the paper with a brief summary 
and outlook.

\section{ Theoretical Background on BAM}
~~While originally derived from a different approach \cite{carne_5}, Eq. (\ref{ba_scaling}) can also be obtained by starting from the kinetic equation 
\cite{trizac1_5, mid1_5, trizac2_5, paul1_5, pathak1_5, pathak2_5}
\begin{equation}\label{kinetic_eq}
 \frac{dn}{dt} = -~{\mbox{``collision cross-section''}} \times v_{\rm{rms}} \times n^2,
\end{equation}
where $n$ is the particle or cluster density and $v_{\rm{rms}}$ is the root-mean-squared velocity of the particles. The collision cross-section 
is proportional to $\ell^{d-1}$, where $\ell$, for spherical particles, can be taken to be their average diameter, which scales 
with the average mass as $m^{1/d}$. For uncorrelated velocity one can take \cite{trizac1_5, trizac2_5, pathak2_5} 
\begin{equation}\label{vrms_mass}
 v_{\rm{rms}} \sim m^{-1/2}.
\end{equation} 
The particle density, given that the total 
mass is conserved, scales inversely with the average mass, i.e., 
\begin{equation}
 n \propto \frac{1}{m}.
\end{equation}
Incorporation of these facts in Eq. (\ref{kinetic_eq}) leads to 
\begin{equation}\label{mass_eq}
 \frac{dm}{dt} = m^{\frac{d-2}{2d}}.
\end{equation}
Solution of Eq. (\ref{mass_eq}) provides time dependence of mass in Eq. (\ref{ba_scaling}).
However, a deviation from Eq. (\ref{vrms_mass}),  can invalidate 
the predictions in Eq. (\ref{ba_scaling}). For $v_{\rm{rms}} \sim m^{-z}$, the growth exponent $\zeta$ becomes \cite{paul1_5}
\begin{equation}\label{zeta_eqn}
 \zeta = \frac{d}{1+dz}.
\end{equation}
\par
~~Starting from Eq. (\ref{kinetic_eq}) and without substituting for any mass dependence of $v_{\rm{rms}}$, one writes \cite{trizac1_5}
\begin{equation}\label{mass_rate}
 \frac{dm}{dt} = m^{\frac{d-1}{d}} v_{\rm{rms}}.
\end{equation}
From Eq. (\ref{mass_rate}), using the time dependence of energy from Eq. (\ref{energ_exp}), and that of mass from Eq. (\ref{mass_exp}), 
after considering that $v_{\rm{rms}} \sim E^{1/2}$, one arrives at 
\begin{equation}
 t^{\zeta-1} = t^{\frac{2\zeta d -2\zeta -\theta d}{2d}},
\end{equation}
by discarding pre-factor(s).
Simple power counting provides the hyperscaling relation \cite{trizac1_5} of Eq. (\ref{hyper_scl}). Eq. (\ref{hyper_scl}) 
in $d=1$, $2$ and $3$ reads 
\begin{equation}
 \theta + 2\zeta =2,
\end{equation}

\begin{equation}\label{2d_hyper}
 \theta +\zeta =2,
\end{equation}
and 
\begin{equation}\label{3d_hyper}
 3\theta +2\zeta =6,
\end{equation}
respectively. We intend to verify these equations for the BAM. We have obtained the time dependence of mass in Eq. (\ref{ba_scaling}) 
by considering Eq. (\ref{vrms_mass}). When this is used in Eq. (\ref{hyper_scl}), 
it straightforwardly appears that mass and energy relate to each other inversely. 

\section{ Models and Methods}
~~For both the models, hard spherical particles, mass being uniformly distributed over the volume or area of the objects,
 move freely between collisions \cite{trizac1_5, gold_5}. 
Mass and momentum remain conserved during the collisions. 
For the BAM, even though the size of the new particle increases, its shape is kept unchanged. For example, two initial spheres of masses 
and diameters ($m_i, \sigma_i$) and ($m_j, \sigma_j$), respectively, coalesce to form a single sphere of mass 
\begin{equation}
 m^{\prime} = m_i + m_j,
\end{equation}
with diameter \cite{trizac1_5}
\begin{equation}
 \sigma^{\prime} = (\sigma_i^d+\sigma_j^d)^{1/d}.
\end{equation}
In this shape retaining process, if the new sphere overlaps with any other particle, then this event is treated as another collision 
and same method of update is applied. The position ($\vec{r}^{~\prime}$) of the centre of mass and the 
velocity ($\vec{v}^{~\prime}$) of the new particle can be obtained 
from the conservation equations \cite{bril3_5} 
\begin{equation}
 m^{\prime}\vec{r}^{~\prime} = m_i\vec{r}_i + m_j \vec{r}_j,
\end{equation}
and
\begin{equation}
 m^{\prime}\vec{v}^{~\prime} = m_i\vec{v}_i + m_j \vec{v}_j,
\end{equation}
where $\vec{r}_i$ and $\vec{r}_j$ are the positions and $\vec{v}_i$ and $\vec{v}_j$ are the velocities 
of particles $i$ and $j$, respectively, before the collision. 
\par
~~In the case of the GGM, the particle velocities are updated via \cite{gold_5, bril3_5}
\begin{equation}\label{velo_ipart}
 \vec{v}_{i}^{~\prime} = \vec{v}_{i} -\Big(\frac{1+e}{2}\Big)[\hat{n}\cdot(\vec{v}_{i}-\vec{v}_{j})]\hat{n}\,,
\end{equation}
and
\begin{equation}\label{velo_jpart}
 \vec{v}_{j}^{~\prime} = \vec{v}_{j} -\Big(\frac{1+e}{2}\Big)[\hat{n}\cdot(\vec{v}_{j}-\vec{v}_{i})]\hat{n}\,,
\end{equation}
where $\vec{v}_i^{~\prime}$ and $\vec{v}_j^{~\prime}$ are the post collisional velocities. In Eqs. (\ref{velo_ipart}) and (\ref{velo_jpart}) 
$\hat{n}$ represents the unit vector parallel to the relative position of the particles $i$ and $j$. In this case, since the colliding 
particles do not undergo coalescence, the particle mass remains unchanged throughout the evolution.
\par
~~We perform event-driven molecular dynamics simulations with these models \cite{rapaport_5, allen_5}, 
where an event is a collision. In this method, since there is no inter-particle 
interaction or external potential, particles move with constant velocities till the next collision. 
Time and partners for the collisions are appropriately identified \cite{allen_5}. 
\par
~~All the results are obtained from simulations in periodic boxes of linear dimension $L$ (equal in all directions), 
in units of the starting particle diameter (see below). Quantitative results are presented after averaging over multiple independent 
initial configurations, the numbers lying between $5$ and $15$.
We start with random initial configurations for both positions and velocities, with $\sigma_i=1$ for all particles. The packing fractions ($\phi$),
values of which will be mentioned 
later, are calculated as $\phi=(N/L^d)\times x$, where $N$ is the initial number of particles in a box, and $x=1$, $\pi/4$ and $\pi/6$ 
in $d=1$, $2$ and $3$, respectively. 
Values of $N$ will also be specified in appropriate places.
In the case of BAM the number of particles decreases with time. So, for this model the number will certainly correspond to the value at the 
beginning of the simulations.
\par
~~For the calculation of the average mass, identification of the clusters is required. For the GGM it was done \cite{paul1_5} by appropriately identifying 
the closed cluster-boundaries within which the packing fraction is higher than a cut-off number $\phi_c$ ($\simeq 0.5$ in $d=1$, $\simeq 0.31$ in $d=2$ and 
$\simeq 0.21$ in $d=3$). 
On the other hand, in the case of BAM, the information on the mass of a cluster is carried by the size of the particles.

\section{ Results}
~~We divide this section into three parts. The first two subsections contain the BAM results from $d=2$ and $d=3$.
 The GGM results are presented in the last one.

\subsection{ BAM in d=2}

\begin{figure}[h!]
\centering
\includegraphics*[width=0.44\textwidth]{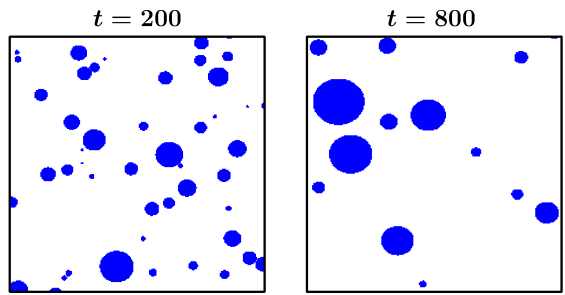}
\caption{\label{fig1} Snapshots during an evolution in the two-dimensional BAM, 
for the packing fraction $\phi=0.08$. The times are mentioned on the top of the frames. The simulation box size is $L=512$. For both 
the times only parts of the original system have been presented.}
\end{figure}

\begin{figure}[htb]
\centering
\includegraphics*[width=0.38\textwidth]{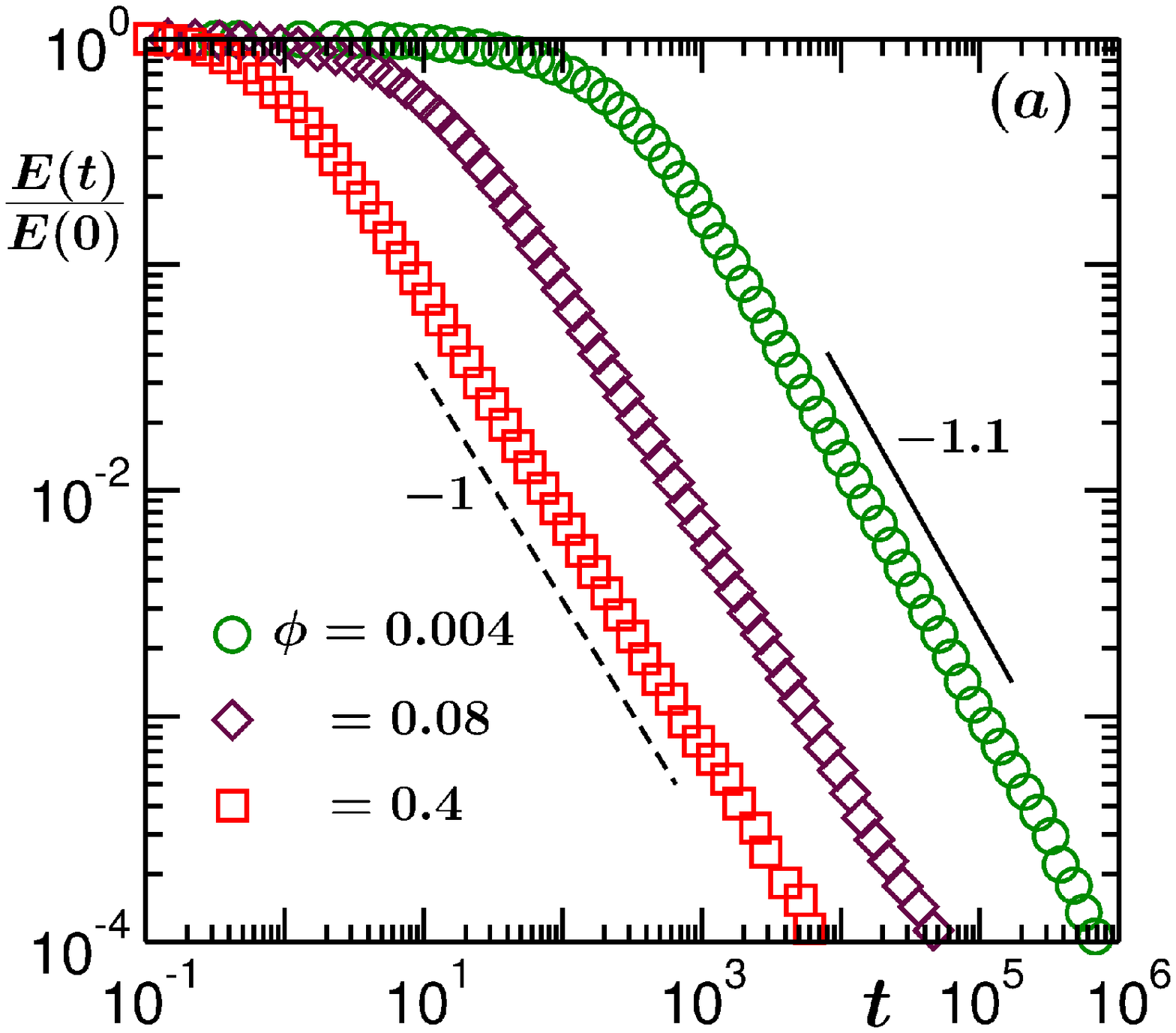}
\vskip 0.75cm
\includegraphics*[width=0.38\textwidth]{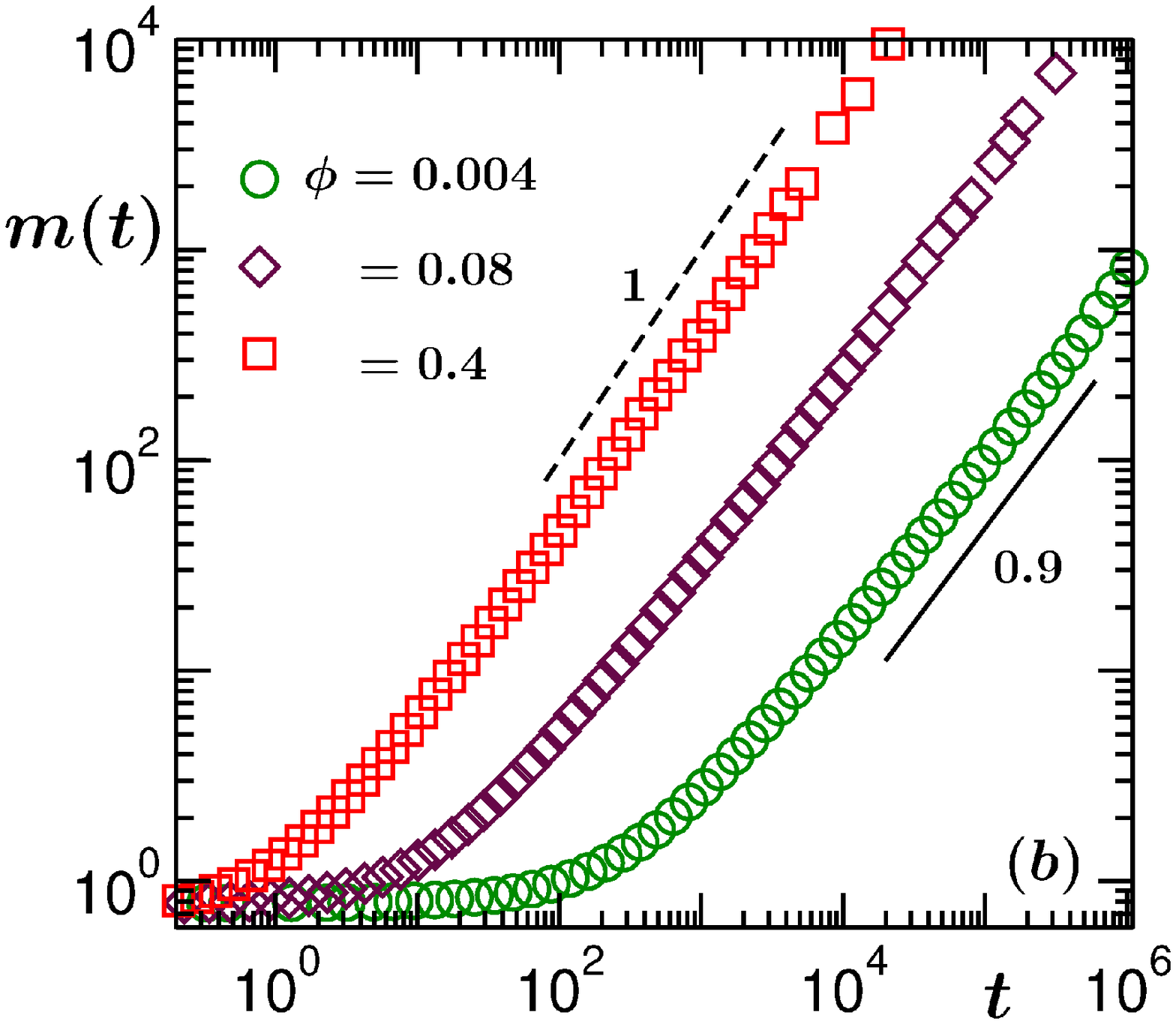}
\caption{\label{fig2} Log-log plots of (a) energy versus time and (b) mass versus time, for different packing fractions (mentioned in the figure). The 
solid and dashed lines represent power-laws. Corresponding exponents are mentioned. All results correspond to the BAM in $d=2$. 
These and other quantitative results in this dimension (for the BAM) are obtained for $N=10^5$.}
\end{figure}

~~In Fig. \ref{fig1} we show two snapshots, obtained during an evolution in the $2D$ BAM. These snapshots 
are from late enough times so that the clusters are reasonably well grown. All the droplets, particularly the smaller ones, 
do not appear perfectly circular. 
This is because of a technical difficulty -- we have divided the whole space to form a discrete lattice system and 
marked the  sites that fall within the boundary of one or the other droplet.
It is clear from these figures that the number of 
clusters is decreasing with time and thus, the average mass of the clusters is increasing.

\begin{figure}[htb]
\centering
\includegraphics*[width=0.375\textwidth]{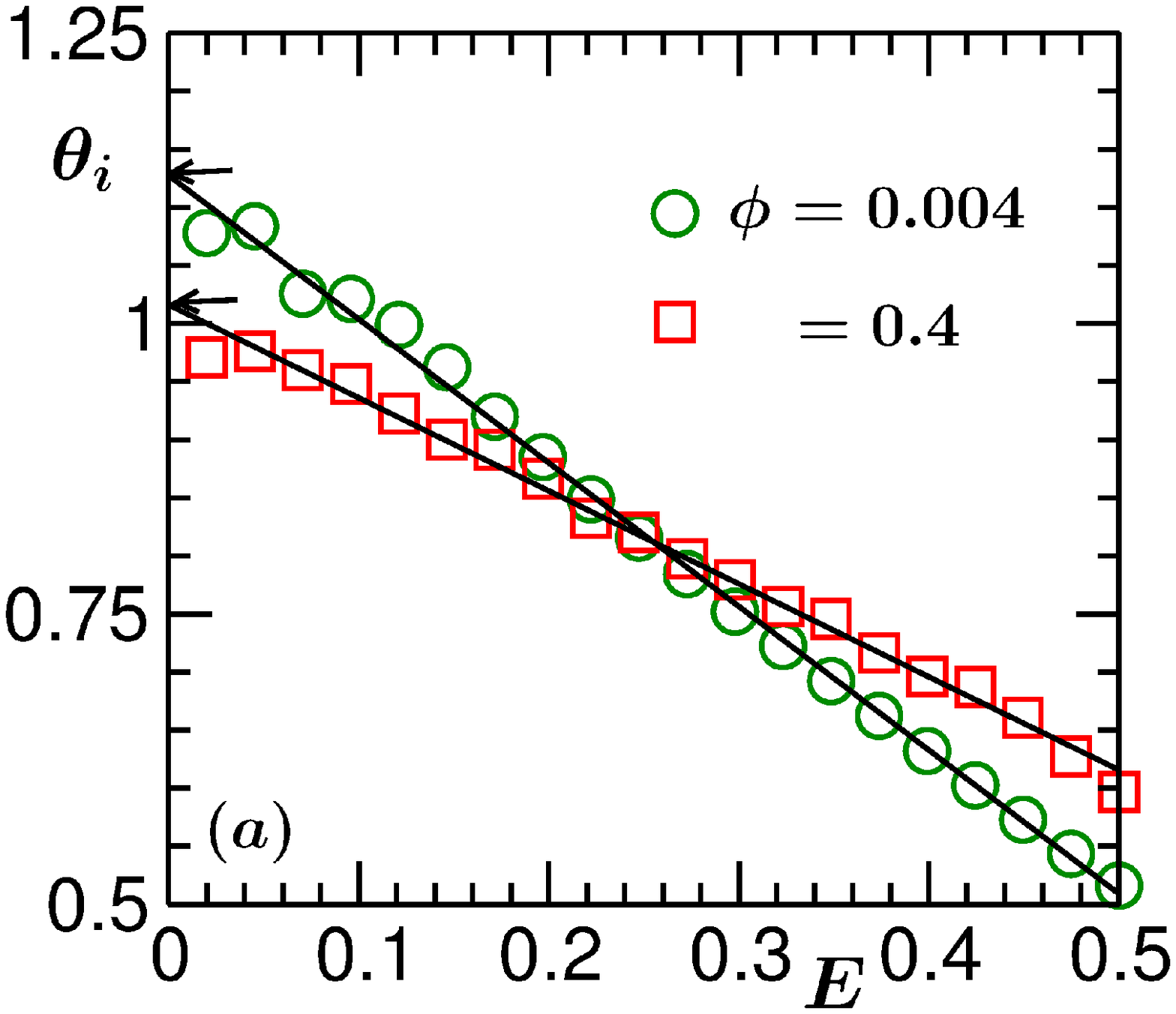}
\vskip 0.75cm
\includegraphics*[width=0.37\textwidth]{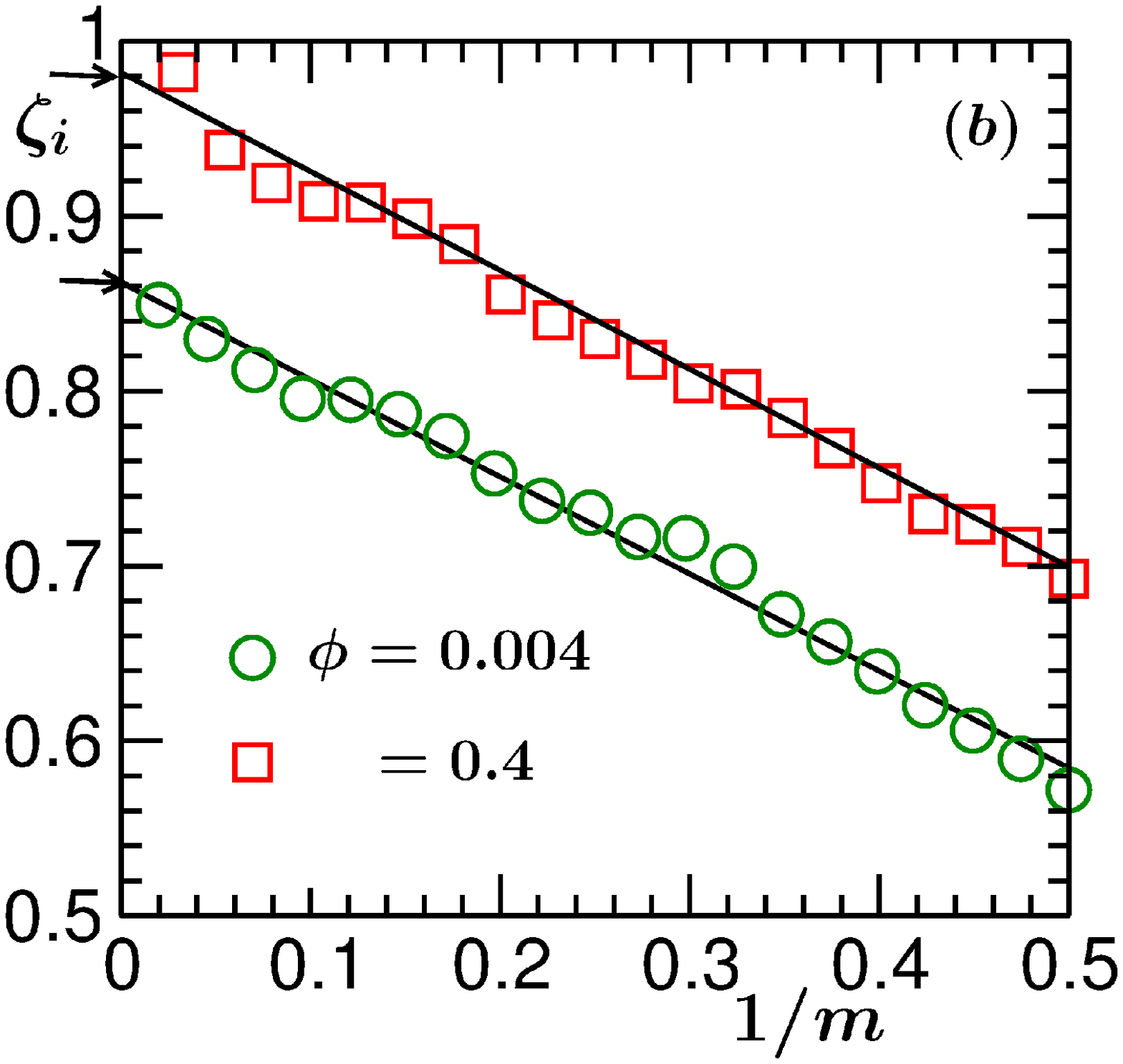}
\caption{\label{fig3} Plots of (a) $\theta_i$ versus $E$ and (b) $\zeta_i$ versus $1/m$, 
 for two values of the packing fraction. 
The results correspond to the BAM in $d=2$. Continuous lines are linear fits to the simulation data sets. The arrows point towards the asymptotic values.}
\end{figure}


\begin{figure}[htb]
\centering
\includegraphics*[width=0.4\textwidth]{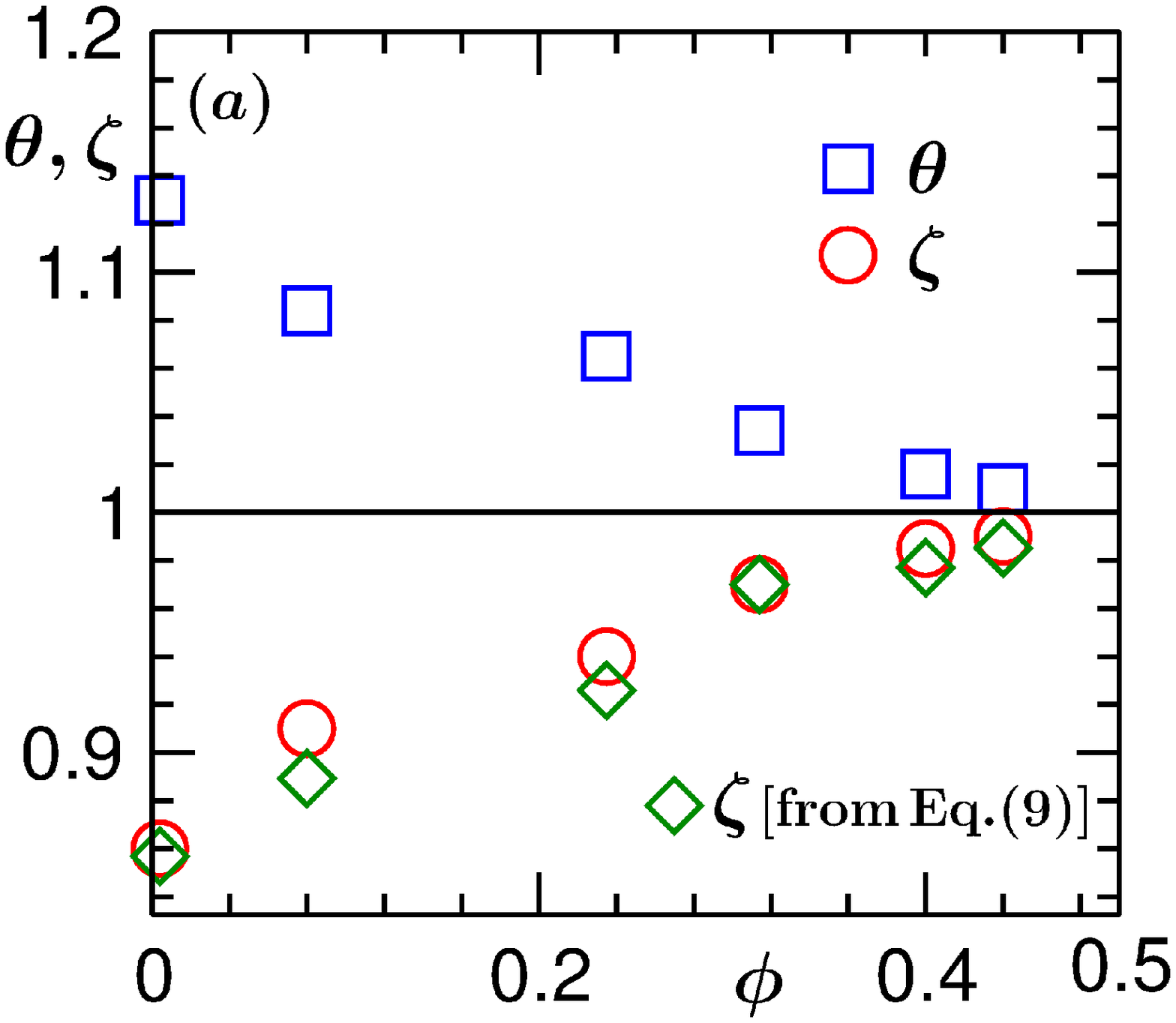}
\vskip 0.75cm
\includegraphics*[width=0.4\textwidth]{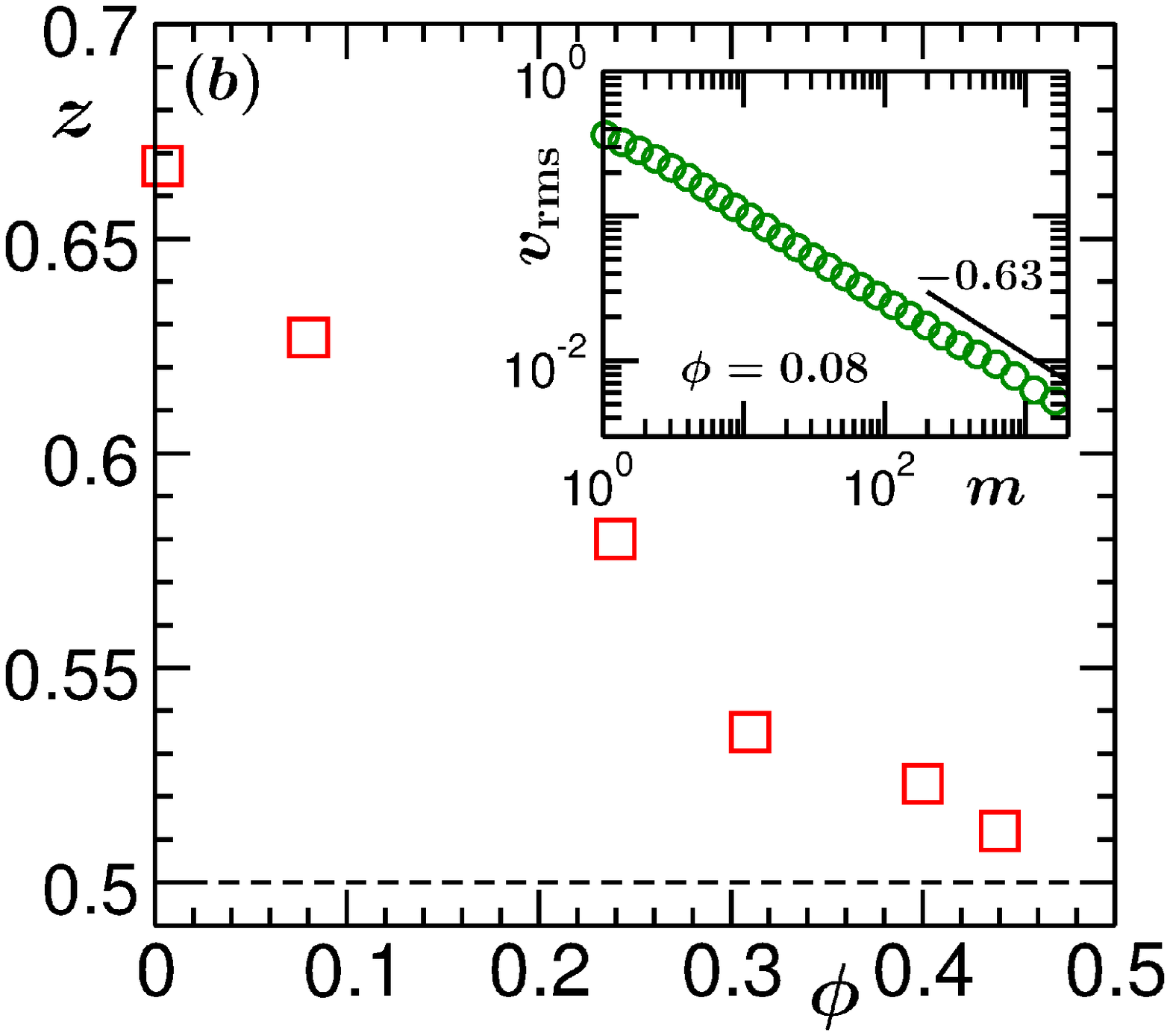}
\caption{\label{fig4} (a) Exponents $\theta$ and $\zeta$ are plotted versus $\phi$. The horizontal line marks the CPY value.
For $\zeta$, we have also included results obtained from Eq. (\ref{zeta_eqn}). (b) Values of the exponent 
$z$ (related to the dependence of $v_{\rm{rms}}$ on $m$) have been plotted versus $\phi$. 
The dashed horizontal line marks the value $0.5$. Inset: Log-log plot of $v_{\rm{rms}}$ versus $m$, for $\phi=0.08$. 
The solid line there corresponds to a power-law, exponent for which has been mentioned. All results are for the $2D$ BAM.}
\end{figure}

\begin{figure}[htb]
\centering
\includegraphics*[width=0.4\textwidth]{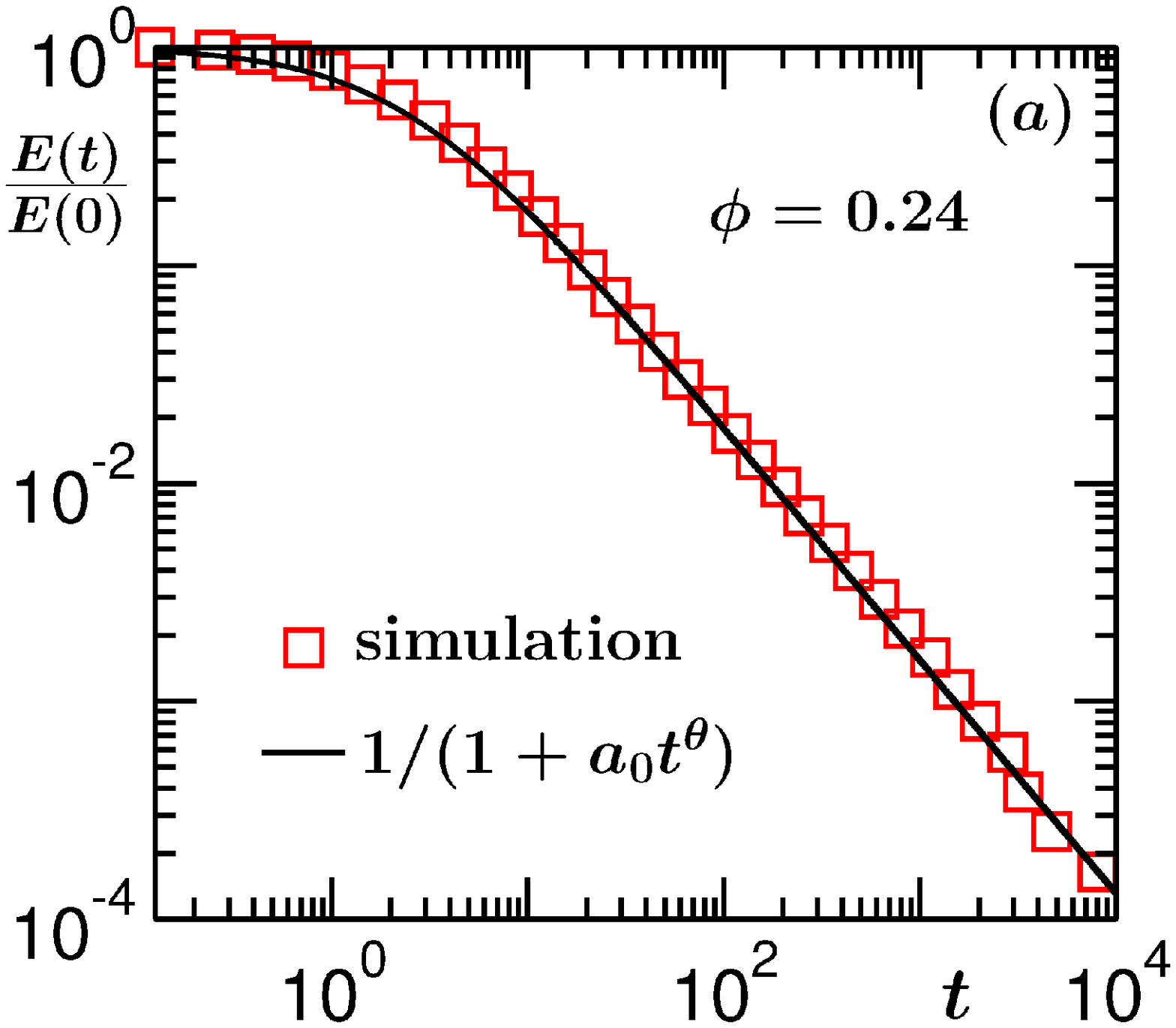}
\vskip 0.75cm
\includegraphics*[width=0.4\textwidth]{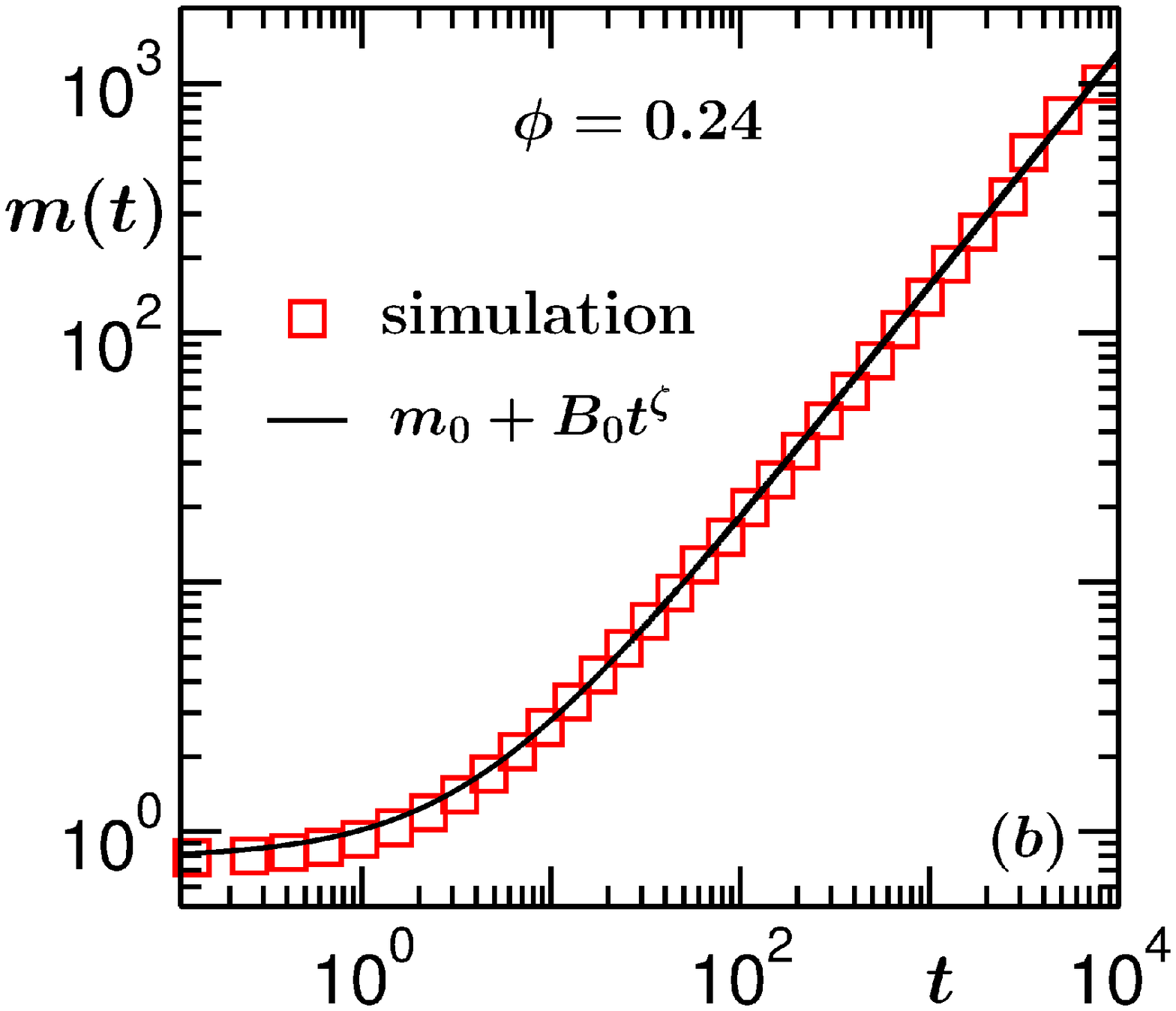}
\caption{\label{fig5} (a) Energy is plotted versus time, on a log-log scale, for the $2D$ BAM with $\phi=0.24$. The solid line 
corresponds to a fitting to Eq. (\ref{enrg_correct}). (b) Same as (a) but for $m$. The solid line here represents Eq. (\ref{mass_correct}).}
\end{figure}
~~

\par
~~In Fig. \ref{fig2}(a) we plot the energy (normalized to unity at $t=0$)
for three different packing fractions, viz., $\phi=0.004, 0.08$ and $0.4$, versus time, 
on a log-log scale. 
Fig. \ref{fig2}(b) shows the log-log plot of the growth of mass 
for the same three values of $\phi$.
While the trends in the long time limit are consistent with power-laws, $\theta$ and $\zeta$, 
the corresponding exponents for the energy decay and growth of mass, respectively, for some densities 
differ from each other, as well as from the CPY \cite{carne_5} value $1$ (recall that we are working in $d=2$).
The deviations from the CPY value are quite significant when $\phi$ is small. 
Value of $\zeta$ increases towards unity \cite{trizac1_5} with the increase of $\phi$.
On the other hand, $\theta$ decreases from a higher value, towards unity, for similar change in $\phi$. 
This already provides hint on the validity of the hyper-scaling relation \cite{trizac1_5}. Here note that a conclusion 
on the power-law exponent from log-log plots or simple data fitting exercises can be misleading. This is because of the presence 
of an offset before the data reach the expected scaling regime, as well as due to the unavailability of 
data over many decades (without being affected by the finite size of the systems).
Thus, to accurately quantify the exponents and confirm the validity of Eq. (\ref{hyper_scl}) [Eq. (\ref{2d_hyper}) in $d=2$] 
we need more accurate quantitative analysis.

\begin{table}[h!]
\caption{Values of $\theta$ and $\zeta$ are listed for different packing fractions, for the $2D$ BAM.}\label{tab1_5}
\centering
\begin{tabular}{|c|c|c|c|}
\hline
\hline
~~~~$\phi$~~~~&~~~~~$\theta$~~~~&~~~~~$\zeta$~~~~&~~~~~$\theta + \zeta$~~~~~\\
\hline
~~~~0.004~~~~&~~~~~1.13~~~~&~~~~~0.86~~~~&~~~~~1.99~~~~~\\
~~~~0.08~~~~&~~~~~1.08~~~~&~~~~~0.91~~~~&~~~~~1.99~~~~~\\
~~~~0.24~~~~&~~~~~1.07~~~~&~~~~~0.94~~~~&~~~~~2.01~~~~~\\
~~~~~0.31~~~~&~~~~~1.03~~~~&~~~~~0.97~~~~&~~~~~2.0~~~~~\\
~~~~~0.4~~~~&~~~~~1.01~~~~&~~~~~0.98~~~~&~~~~~1.99~~~~~\\
~~~~~0.44~~~~&~~~~~1.01~~~~&~~~~~0.99~~~~&~~~~~2.0~~~~~\\
\hline
\hline
\end{tabular}
\end{table}

\par
~~For this purpose, we calculate the instantaneous exponent $\theta_i$, for the decay of $E$, defined as \cite{huse_5}
\begin{equation}\label{inst_theta}
 \theta_i= -\frac{d ({\rm{ln}}E)}{d ({\rm{ln}}t)},
\end{equation}
accepting that a power-law behavior indeed exists.
In Fig. \ref{fig3}(a) we plot $\theta_i$ as a function of $E$. 
For the sake of clarity, here we show the plots for $\phi=0.004$ and $0.4$ only.  
In both the cases linear behavior is visible over an extended range.
We extract the asymptotic value, $\theta$, from the convergence of $\theta_i$ in the 
$t \rightarrow \infty$, i.e.,  $E \rightarrow 0$ limit. Indeed, $\theta$ exhibits density dependence.
\par
~~Similar exercise has also been performed for the growth of mass. 
In Fig. \ref{fig3}(b) we plot the instantaneous exponent  $\zeta_i$, for the growth of mass, defined as \cite{huse_5}
\begin{equation}\label{inst_zeta}
 \zeta_i = \frac{d ({\rm{ln}}m)}{d ({\rm{ln}}t)},
\end{equation}
as a function of $1/m$, for $\phi=0.004$ and $0.4$. Here also we obtain asymptotic values from linear extrapolations.
Clearly, the numbers vary with the change in $\phi$.
The exponents  
 $\theta$ and $\zeta$, obtained from these exercises, for different values of $\phi$, are quoted in Table \ref{tab1_5}. 

\par
~~ CPY \cite{carne_5} predict that the energy decay and the growth of mass are inversely proportional to 
each other, with $\theta = \zeta =1$. Our results show that the exponents, which have been 
accurately quantified via the calculation of instantaneous exponents \cite{huse_5},  are 
nonuniversal, with strong dependence upon the packing fraction. 
We observe that the CPY predictions tend to be valid only at higher values of $\phi$. 
For lower values of $\phi$ they deviate significantly.
But the simulation results follow the relation \cite{trizac1_5}:  $\theta+\zeta =2$, to a good accuracy -- 
see the numbers quoted in the last column of Table \ref{tab1_5}.  While the numbers in Table \ref{tab1_5} provide 
accurate information, to get a feel about how the convergence towards the CPY exponent occurs, in Fig. \ref{fig4}(a)
 we show plots of $\theta$ and $\zeta$, versus $\phi$.
\par
~~In Fig. \ref{fig4}(a) we have also presented data for $\zeta$ which were estimated via Eq. (\ref{zeta_eqn}). This data set shows similar 
trend as the one obtained via the calculation of $\zeta_i$. To apply Eq. (\ref{zeta_eqn}), we have estimated $z$ by calculating $v_{\rm{rms}}$ 
at different times. A plot of $z$ as a function of $\phi$ is shown in Fig. \ref{fig4}(b).
In the inset of Fig. \ref{fig4}(b) we presented a log-log plot of $v_{\rm{rms}}$ vs $m$, for $\phi=0.08$. The solid line there, consistent with the 
simulation data, 
represents a power-law with exponent $z=0.63$, that differs significantly from $0.5$ that is needed to validate the prediction of CPY. Here note that 
$z$ was estimated via fitting of the data in the inset to a power-law form.
From the main frame of Fig. \ref{fig4}(b) we notice that $z$ reaches $0.5$ approximately when $\phi=0.45$.
\par
~~While these results of ours are consistent with previous 
reports \cite{trizac1_5}, such accurate analyses are new. On the other hand, in $d=3$ simulation study to confirm the validity of the 
hyperscaling relation was not performed earlier, to the best of our knowledge. In the next subsection we present these results. 

\par
~~Before moving to the next subsection, we provide further discussions on the $d=2$ results which may be valid in $d=3$ as well. 
We have accepted linear behavior of the data sets in Fig. \ref{fig3}, for energy as well as mass. Given the statistical fluctuation in 
the presented results, further checks of this assumption is necessary. Moreover, what scaling forms such linear trends imply?
\par
~~For a linear behavior of the $\theta_i$ versus $E$ data, one can use 
\begin{equation}\label{theta_inst}
 \theta_i = \theta -AE,
\end{equation}
in the definition in Eq. (\ref{inst_theta}), to write 
\begin{equation}\label{e_t_new}
 \frac{dE}{AE^2-\theta E} = \frac{dt}{t},
\end{equation}
where $A$ is the slope of a $\theta_i$ versus $E$ plot and $AE < \theta$. Then Eq. (\ref{e_t_new}) provides
\begin{equation}\label{enrg_correct}
 E = \frac{\theta/A}{1+a_0 t^{\theta}}\,,
\end{equation}
where $a_0$ is a positive constant. 
This implies, the value of 
$E$ at $t=0$ provides a non-zero slope in Fig. \ref{fig3}(a) and this off-set is also responsible for the misleading trend of $E$ versus $t$ data
on a double-log scale, over early decades. 
However, this single scaling form will be completely true if a linear behavior in Fig. \ref{fig3}(a) 
is realized from $t=0$. This, in fact, is not the case. For $E>0.5$ there exists slight bending (data not shown). This implies correction to the 
form in Eq. (\ref{enrg_correct}). Furthermore, had there been no correction, the data sets in Fig. \ref{fig3}(a) would have been described by 
\begin{equation}\label{theta_eqn}
 \theta_i = \theta (1-E),
\end{equation}
implying same values for the $y$- intercept and the slope, i.e.,
\begin{equation}\label{enrg_slope}
 A = \theta.
\end{equation}
This fact, in absence of a correction, automatically leads to the initial condition $E=1$ at $t=0$. 
Here recall that everywhere we have normalized $E$ by its value at $t=0$.
Eq. (\ref{theta_eqn}) can also be checked by using 
Eqs. (\ref{enrg_correct}) and (\ref{enrg_slope}) 
in Eq. (\ref{inst_theta}). However, in reality small disagreement exists between $A$ and $\theta$, 
when we fit the data sets in Fig. \ref{fig3}(a) to the form in Eq. (\ref{theta_inst}).
\par
~~In Fig. \ref{fig5}(a) we have shown a comparison between the simulation data and fit to the mathematical form in Eq. (\ref{enrg_correct}), for $\phi=0.24$, 
by fixing the corresponding value of $\theta$ to the number mentioned in Table \ref{tab1_5} and asserting that $\theta =A$. 
A near perfect agreement is observed. This substantiates the linear 
assumption in Fig. \ref{fig3}(a), as well as confirms the absence of any strong correction in the early time decay. This is consistent with the fact that $\theta/A$ 
differs from unity by approximately $10\%$.
\par
~~Similarly,  considering the 
linear trend in Fig. \ref{fig3}(b), one obtains 
\begin{equation}\label{mass_correct}
 m = m_0 +B_0 t^{\zeta},
\end{equation}
where $B_0$ is a constant amplitude and $m_0$ is the average initial mass. In Fig. \ref{fig5}(b) we show an exercise,  analogous to 
Fig. \ref{fig5}(a), by fitting simulation data for mass to Eq. (\ref{mass_correct}). 
Here the continuous line is obtained by fixing $m_0$ to $\pi/4$ (which indeed is the starting mass),
 $\zeta$ to the value quoted in 
Table \ref{tab1_5} corresponding to $\phi=0.24$ and using $B_0$ as an adjustable parameter. 
Once again, the agreement is nice, validating the linear assumption and 
discarding any possibility of a strong correction. 
Here we mention that in the 
literature of growth kinetics, such linear trends in the time-dependent exponents have been mis-interpreted as strong 
corrections to scaling -- see Refs. \cite{amar_5, majum1_5} for discussion.

\subsection{ BAM in d=3}
\begin{figure}[h!]
\centering
\includegraphics*[width=0.36\textwidth]{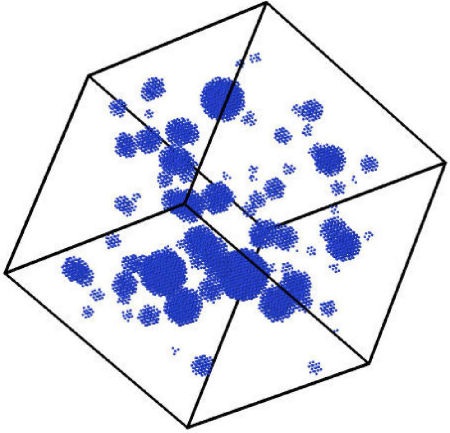}
\caption{\label{fig6} An evolution snapshot for the $3D$ BAM, from $t=100$. The  packing fraction and the  linear 
dimension of the cubic box are $0.052$ and $64$, respectively.}
\end{figure}

\begin{figure}[htb]
\centering
\includegraphics*[width=0.39\textwidth]{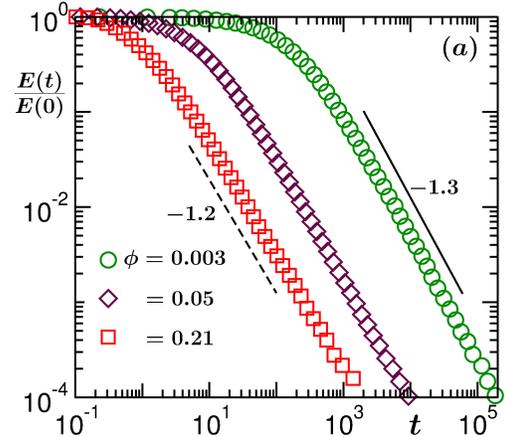}
\vskip 0.75cm
\includegraphics*[width=0.42\textwidth]{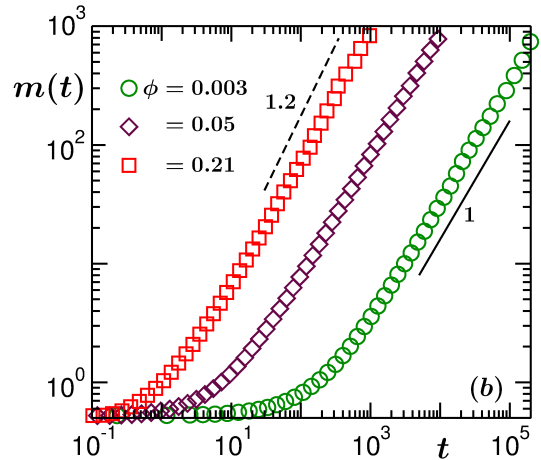}
\caption{\label{fig7} Log-log plots of (a) energy versus time and (b) mass versus time,
 for the $3D$ BAM. Results from three different packing fractions are included. 
The solid and dashed lines are power-laws, exponents for which are mentioned. In this dimension, like in the $2D$ case,
 all the quantitative results for the BAM are obtained from simulations (in cubic boxes) with $N=10^5$.}
\end{figure}

\begin{figure}[htb]
\centering
\includegraphics*[width=0.39\textwidth]{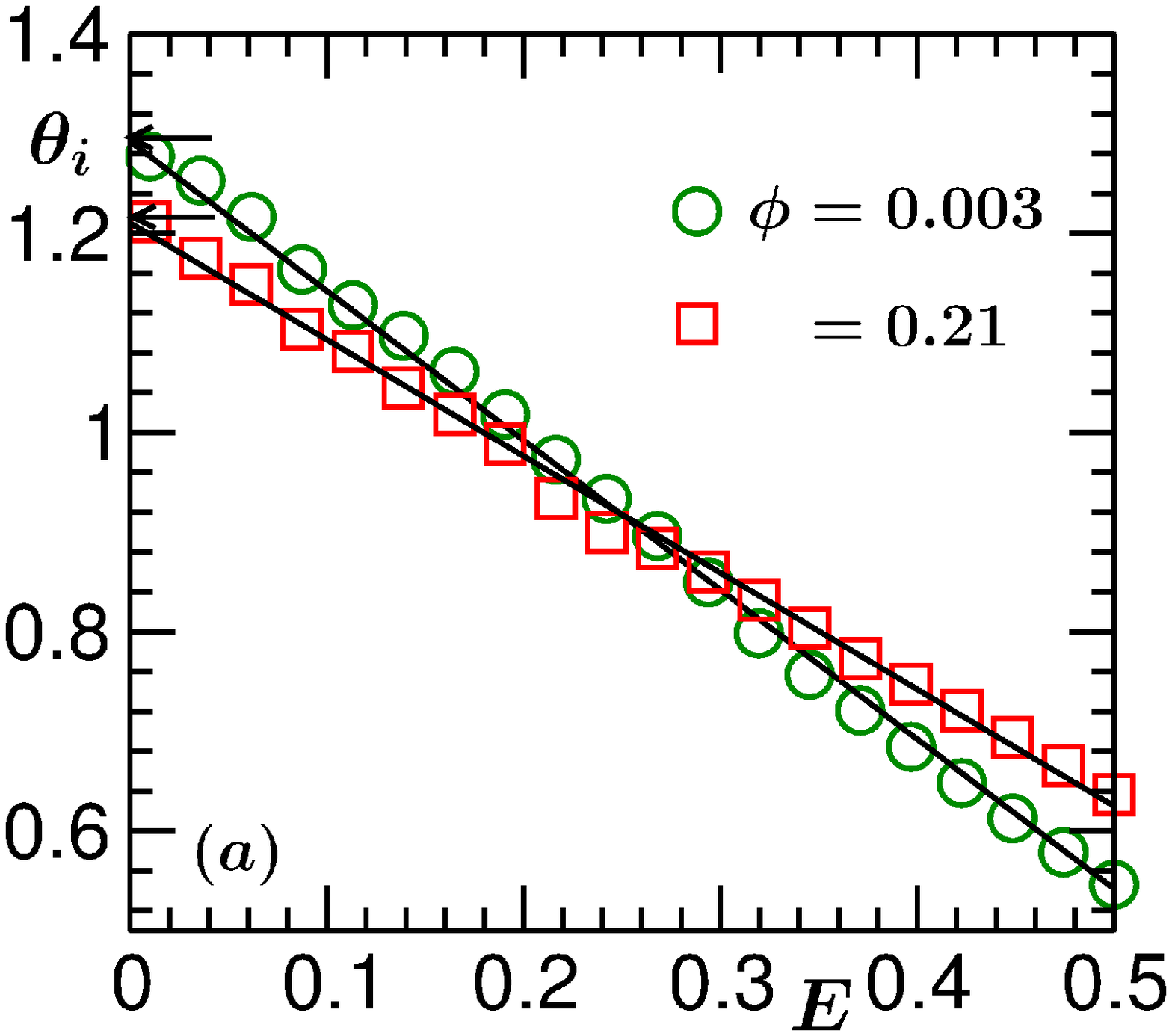}
\vskip 0.75cm
\includegraphics*[width=0.40\textwidth]{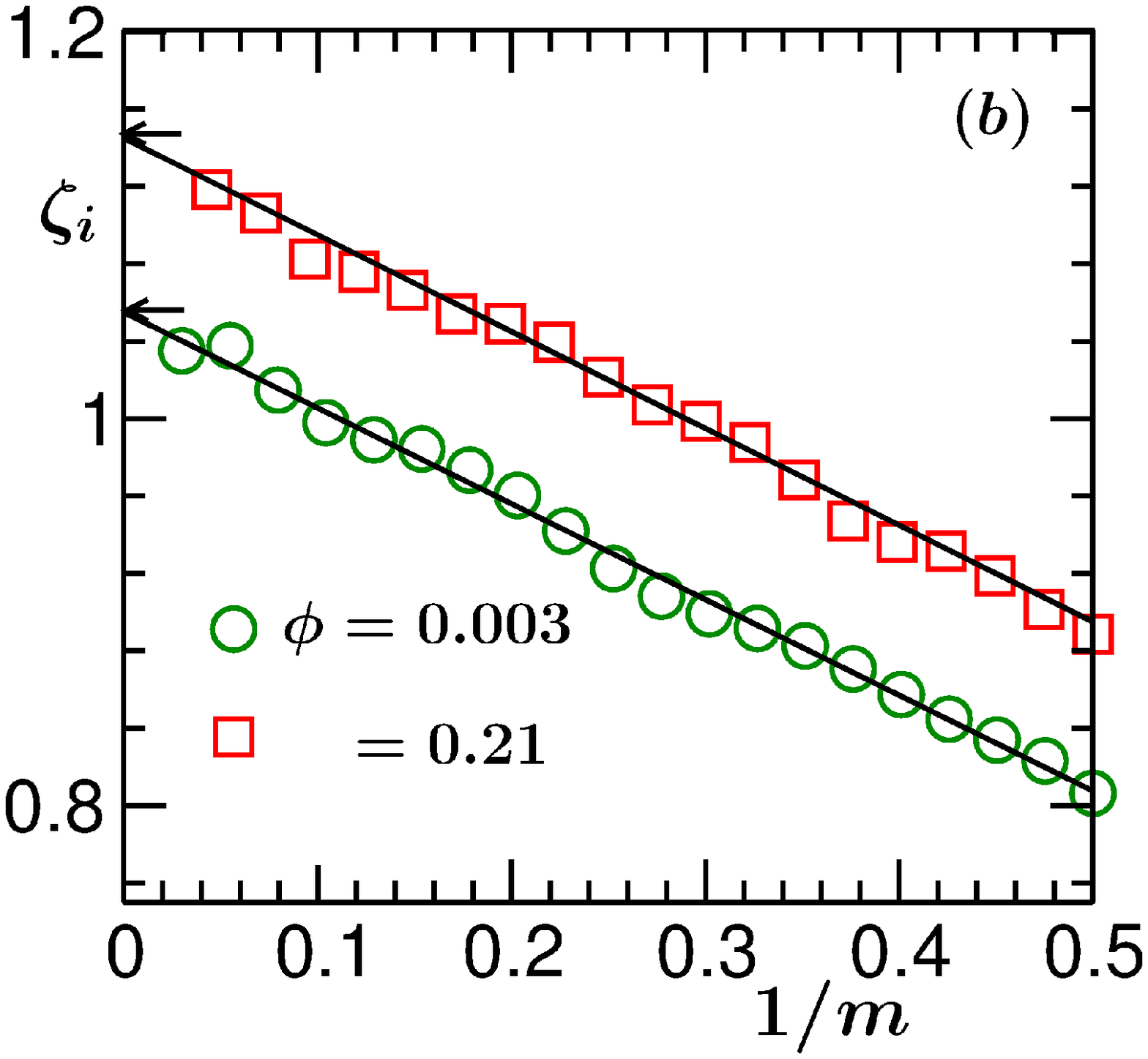}
\caption{\label{fig8} Plots of the instantaneous exponents (a) $\theta_i$ and (b) $\zeta_i$, versus $E$ and $1/m$, respectively, 
 for the $3D$ BAM. The solid straight lines are linear fits to the simulation data sets. The arrows mark the asymptotic values. 
We have shown results from two values of $\phi$.}
\end{figure}


\begin{figure}[htb]
\centering
\includegraphics*[width=0.4\textwidth]{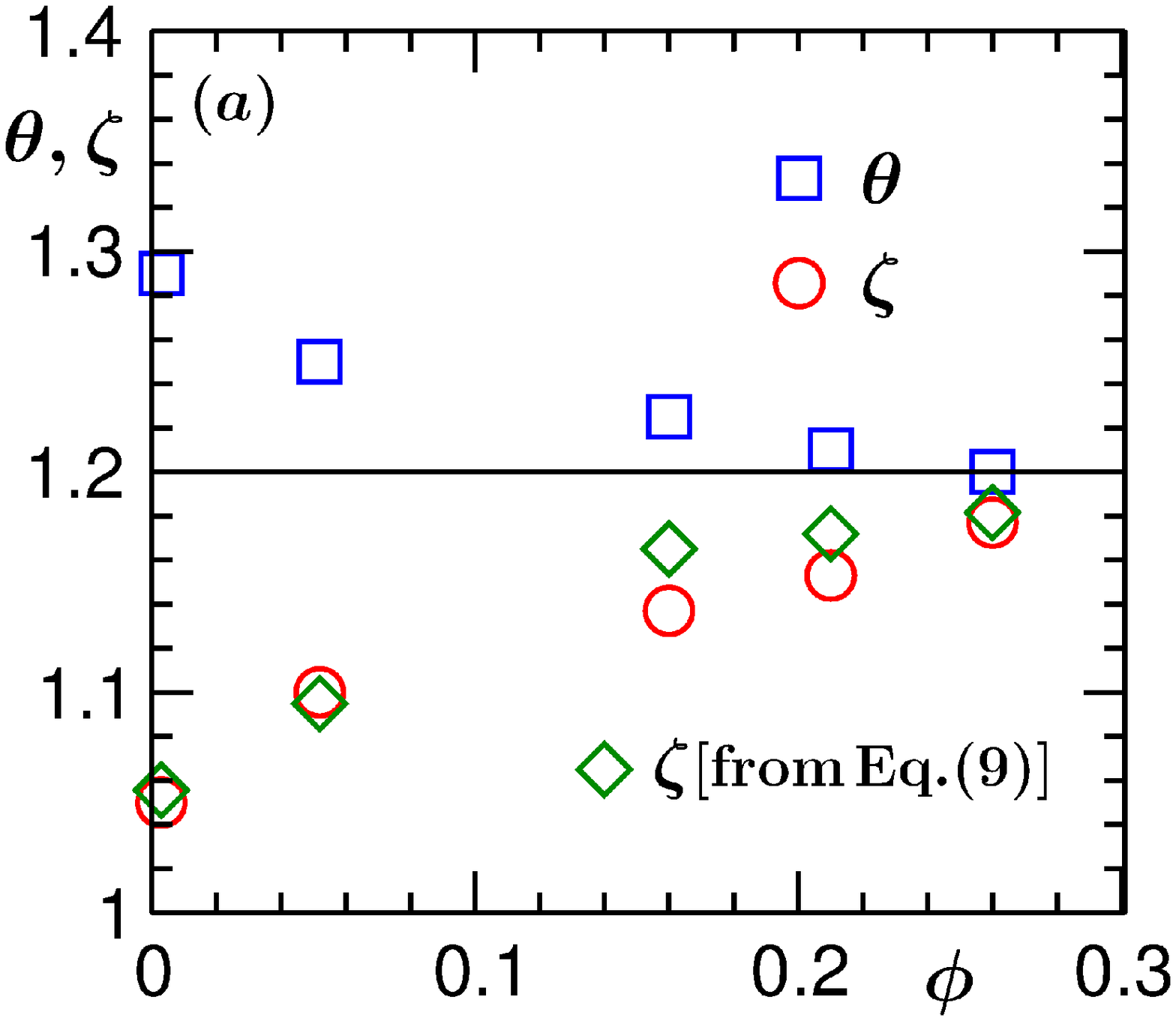}
\vskip 0.75cm
\includegraphics*[width=0.39\textwidth]{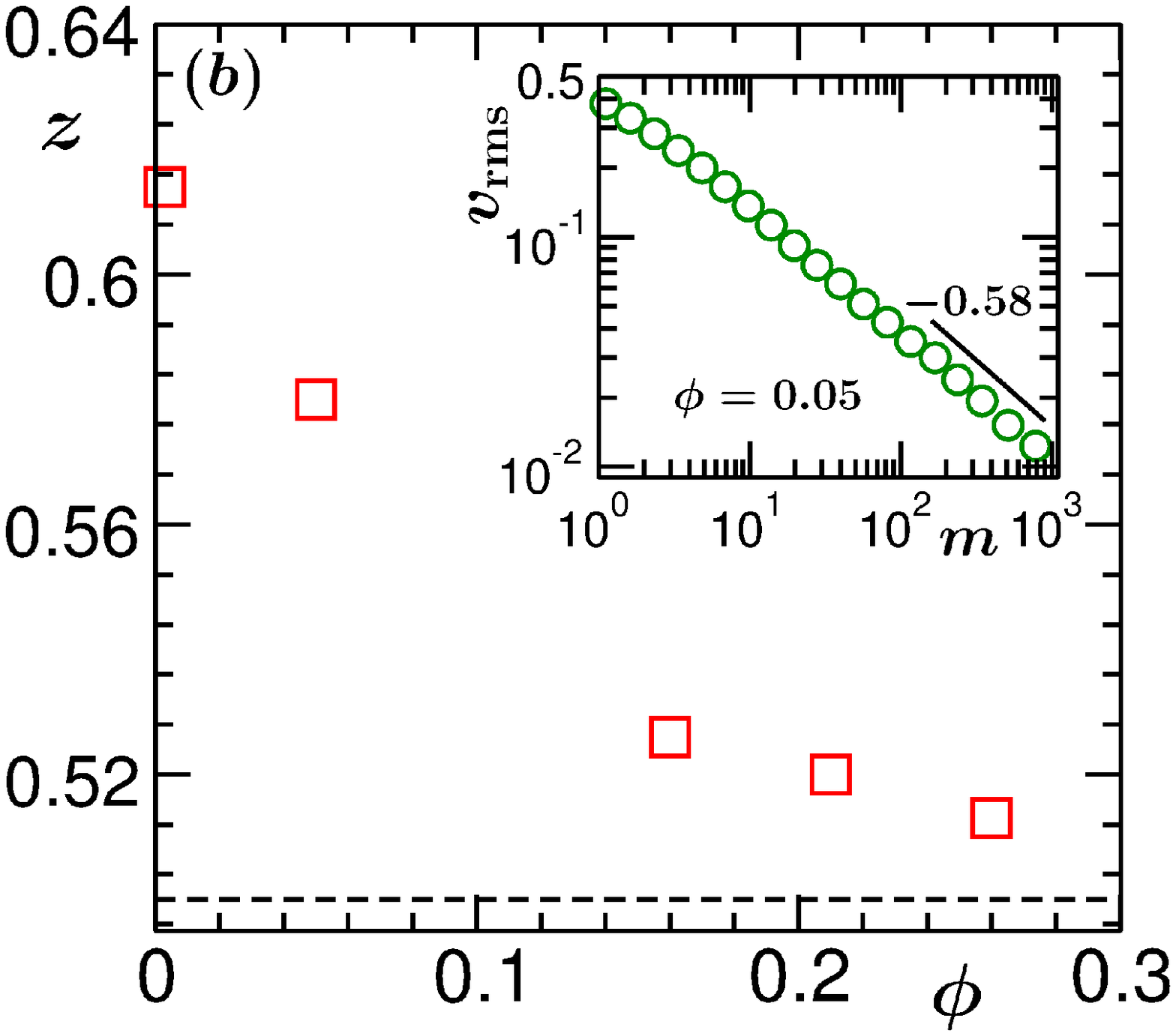}
\caption{\label{fig9} (a) Exponents $\theta$ and $\zeta$ are plotted versus $\phi$. The horizontal line represents the CPY value.
 Values of $\zeta$ calculated from Eq. (\ref{zeta_eqn}) are also included. (b) Exponent 
$z$ is plotted versus $\phi$. 
Dashed horizontal line there corresponds the value $0.5$. 
Inset shows a log-log  plot of $v_{\rm{rms}}$ versus $m$, for $\phi=0.05$. 
The solid line there corresponds to a power-law,  the exponent for which is mentioned. 
All results correspond to the $3D$ BAM.}
\end{figure}

~~Given that the context is same and primary discussions have been provided in the previous subsection, here we straightway present the results.
First, in Fig. \ref{fig6} we show a snapshot for the $3D$ BAM evolution. Like in $d=2$, here also lesser sphericity is visible for 
smaller particles. This is because of the technical reason mentioned in the previous subsection.
\par
~~In Fig. \ref{fig7}(a) we plot the energy as a function of time, on a log-log scale,  
for various different choices of the packing fraction. Prediction of CPY \cite{carne_5} for the exponent 
for the energy decay, as well as that for the growth of mass, is 
$6/5$, in this space dimension. The values of  $\theta$, as can be judged from Fig. \ref{fig7}(a), do not obey this  theoretical 
number for all values of $\phi$, like in $d=2$. In this dimension also  $\theta$ seems to be decreasing from a higher value towards $6/5$, 
as the packing fraction increases. In Fig. \ref{fig7}(b) we show log-log plots of average mass of the clusters as a function of time, for 
the same choices of the  packing fraction.
Unlike the energy decay, here the value of the exponent $\zeta$ increases
towards the value $6/5$ with the increase of $\phi$. This fact is also similar to the case of $d=2$.

\begin{table}
\caption{Values of $\theta$ and $\zeta$ for different packing fractions. All results are for $3D$ BAM.}\label{tab2_5}
\centering
\begin{tabular}{|c|c|c|c|}
\hline
\hline
~~$\phi$~~&~~~$\theta$~~&~~~$\zeta$~~&~~$3\theta + 2\zeta$~~\\
\hline
~~0.003~~&~~~1.29~~&~~~1.05~~~&~~~5.97~~\\
~~0.05~~&~~~1.25~~ &~~~1.10~~&~~~5.95~~\\
~~0.16~~&~~~1.22~~ &~~~1.14~~&~~~5.94~~\\
~~~0.21~~&~~~1.21~~&~~~1.15~~&~~~5.93~~\\
~~~0.26~~&~~~1.2~~&~~~1.17~~&~~~5.94~~\\
\hline
\hline
\end{tabular}
\end{table}

\par
~~For more accurate quantification of the exponents, for the energy decay as well as for the growth of mass, we calculate the 
instantaneous exponents  \cite{huse_5} $\theta_i$ and $\zeta_i$, defined earlier, 
and plot them versus $E$ and $1/m$, respectively,  in Figs. \ref{fig8}(a) and  \ref{fig8}(b), for $\phi=0.003$ and $0.21$. 
The asymptotic values, estimated from these plots of instantaneous 
exponents, by assuming linear behavior of the data sets, are quoted in Table \ref{tab2_5}. 
It can be observed that, like in $d=2$, the exponents are strongly $\phi$-dependent. However, they obey the 
relation \cite{trizac1_5}  $3\theta + 2\zeta =6$, within $1\%$ deviation. Again, for a visual feeling, in Fig. \ref{fig9}(a) we show $\theta$ 
and $\zeta$ with the variation of $\phi$. 
\par
~~In this dimension also, for $\zeta$, we have shown results from calculations  via Eq. (\ref{zeta_eqn}). Again, trends of the data sets, 
obtained via convergence of $\zeta_i$ and from Eq. (\ref{zeta_eqn}), are very similar. Discrepancies that are observed here and in the previous 
subsection can be attributed to the fact that even though the values of $z$ in both the dimensions are obtained from 
exercises involving 
the instantaneous exponent, quality of data is rather poor. 
A plot of $z$ vs $\phi$ is shown in Fig. \ref{fig9}(b). The inset of this figure demonstrates the 
consistency of the $v_{\rm{rms}}$ vs $m$ data with the estimated exponent, for $\phi=0.05$. 
\par
~~Thus, the hyperscaling relation [Eq. (\ref{3d_hyper})] is valid, to a good accuracy, for all values of $\phi$ and 
the CPY exponent appears reasonably accurate only for high packing fraction. 
The critical numbers turn out to be $\phi \gtrsim 0.45$ and $\gtrsim 0.25$, respectively, in $d=2$ and $3$, by considering $2\%$ deviation from the 
CPY value as acceptable.
Here note that in $d=1$, CPY and, thus, the hyperscaling relation are observed to be valid for all previously studied densities \cite{carne_5, paul1_5}.
\par
~~Like in the $2D$ case, $\theta_i$ vs. $E$ curves show nice linear trend in $d=3$ also, for $E \leq 0.5$. This implies the validity of Eq. (\ref{enrg_correct}) 
over most part of the energy decay. Similar conclusion applies to the case of  mass. We have indeed checked the accuracy of Eqs. (\ref{enrg_correct}) 
and (\ref{mass_correct}) by 
comparing them with the simulation data for $E$ versus $t$ and $m$ versus $t$. Excellent agreements have been observed. However, for the sake of 
brevity we do not present these results.

\subsection{ The case of GGM}

\begin{figure}[htb]
\centering
\includegraphics*[width=0.32\textwidth]{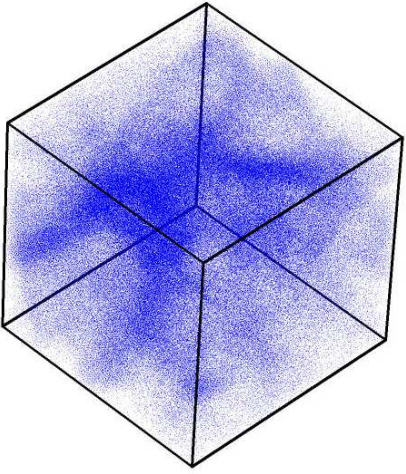}
\caption{\label{fig10}An evolution snapshot for the $3D$ GGM with $e=0.8$. The packing fraction is $0.1$ and 
the linear dimension of the simulation box is $L=120$. Locations of the particles are marked.}
\end{figure}

\begin{figure}[h!]
\centering
\includegraphics*[width=0.4\textwidth]{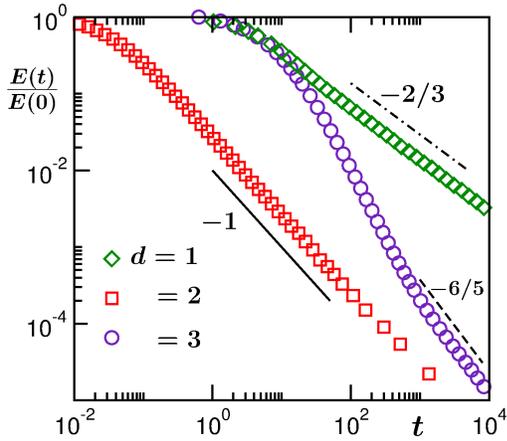}
\caption{\label{fig11} Log-log plots of the decay of energy in the GGM in $d=1$, $2$ and $3$. The dashed-dotted, dashed and solid lines correspond 
to power-laws, exponents for which are mentioned next to them. The presented results correspond to 
 $\phi=0.3$, $0.3$, $0.1$ and $L=32768$, $512$, $120$ for $d=1$, $2$ and $3$, respectively. The data sets have been multiplied by appropriate 
factors to bring them within the presented abscissa and ordinate scales.}
\end{figure}

\begin{figure}[h!]
\centering
\includegraphics*[width=0.44\textwidth]{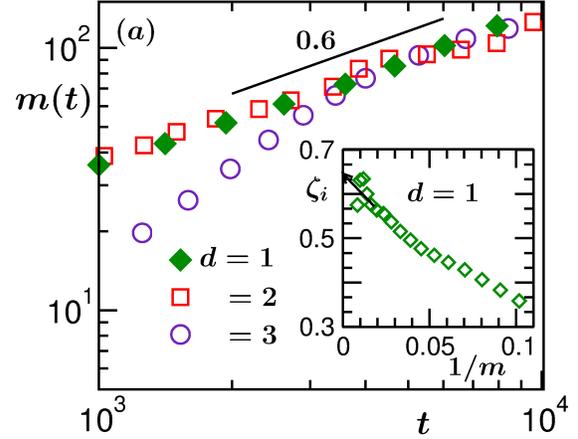}
\vskip 0.75cm
\includegraphics*[width=0.44\textwidth]{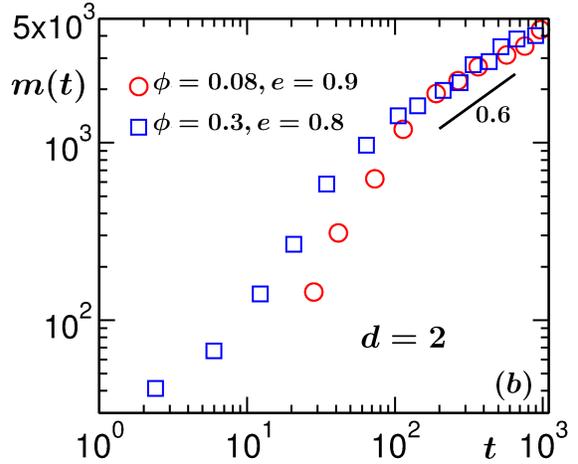}
\caption{\label{fig12} (a) Log-log plots of the growth of the average mass in  all the three dimensions, for the GGM. The data sets 
have been multiplied by appropriate numbers 
to bring them onto the scales of the graph. The solid line corresponds to a power-law with an exponent $0.6$. 
The system sizes and packing fractions for different dimensions are same as in Fig. \ref{fig11}.
The discrepancy of $d=3$ data at early time is related to longer crossover time to ICS, for the chosen coefficient of restitution and 
density of particles. Inset: Plot of instantaneous exponent, $\zeta_i$, vs $1/m$, for the $1D$ GGM. The solid line there is a guide to the eye.
(b) Log-log plots of the growth of average mass for different combinations of coefficient of restitution and 
packing fraction, in $d=2$. 
The data set corresponding to $\phi=0.08$ has been  appropriately scaled to bring it within the presented ranges of the axes. 
The solid line there is a power-law with an exponent $0.6$.}
\end{figure}

~~As stated earlier, even though the particles do not stick to each other, inelastic collisions lead to clustering in the GGM.
However, unlike the case of BAM, in this case, over an initial period of time, referred to as the homogeneous cooling state (HCS) \cite{gold_5},
 density in the system remains 
uniform. After a certain time, value of which depends upon the overall density of particles and the choice of $e$, the system falls unstable to 
fluctuations and crosses over to a clustering regime, referred to as the inhomogeneous cooling state (ICS) \cite{gold_5}. 
In the HCS the energy decay follows the Haff's law \cite{haff_5}
\begin{equation}
 E = (1+ct)^{-2}\,,
\end{equation}
where $c$ is a dimension dependent constant. While the decay in HCS, apart from $c$, is dimension independent, it has been established that 
the exponent in ICS is strongly dimension dependent \cite{nie_5, paul1_5, pathak1_5, naim3_5, shinde1_5}. 
On the other hand, no appropriate conclusion has been drawn \cite{paul1_5, das1_5, das2_5, paul2_5, luding_5} with respect to the 
dimension dependence of the growth of the average mass of clusters.
In this subsection, while the primary 
objective is to investigate the latter issue in the GGM, we present results for the decay of energy also. For both the quantities
 our focus will be on ICS. 

~~We start by showing a representative snapshot, in Fig. \ref{fig10}, from the evolution in GGM in $d=3$. The snapshot shows high and low density 
regions, like the phase separation during a vapor-liquid (VL) transition \cite{roy1}. 
The morphology here is interconnected, that resembles the 
ones for high overall density (close to the critical value) in the VL transitions \cite{roy1}. 
Nevertheless, there exist differences. The equal-time correlation function \cite{bray_5}, that provides 
quantitative information on pattern formation, does not \cite{das1_5, das2_5} 
exhibit intermediate distance oscillation (around zero) in GGM as strong as that for the VL transition \cite{roy1, das3_5}. 
A reason behind such structural difference is that, while in the VL transition phase separation is driven by inter-particle interaction, the clustering 
in the GGM is related to the velocity parallelization due to inelastic collisions. That way, the structure, and thus, the correlation function 
for the GGM, should have more similarity with that for the active matter systems where the direction of motion of a particle is strongly 
influenced by the average direction of motion of the neighbors, e.g., in the Vicsek model \cite{vicsek_5, das4_5}. 
In any case, given the structural difference between the GGM and BAM, a similarity in the dynamics is not really expected. Below we substantiate this.
\par
~~In Fig. \ref{fig11} we show plots for the decay of energy, from $d=1$, $2$ and $3$, for the GGM. Note that the axes are scaled to bring all the 
plots within appropriate abscissa and ordinate ranges that can help make the crucial features identifiable for all values of $d$.
Clearly, the decay rate at late time (in the ICS) is different 
for different dimension. Interestingly, the exponents are in nice agreement with $2d/d+2$, predicted by CPY \cite{carne_5} -- see the 
consistency of the data sets with various power-law lines.
At much later time (not shown) the decays are faster, which can be related to finite-size effects. 
The presented results are consistent with other simulation studies \cite{paul1_5, shinde1_5, naim3_5, nie_5}.
On the other hand, from some previous studies on growth of mass \cite{paul1_5, paul2_5}, 
we got hint that this agreement of energy decay with the prediction of 
 CPY \cite{carne_5} may be accidental and should have different reason. 
To make a more concrete statement on this aspect, below we look at the growth picture. 
\par
~~In Fig. \ref{fig12}(a) we present plots of $m$ versus $t$, on a log-log scale, for all the three dimensions. 
We discard data affected by the finite size of the systems. 
Furthermore, like in the plots of energy decay, data from different dimensions have been multiplied by different factors. 
It appears, all the data sets exhibit power-laws, at late times, with very similar exponent, which is close to $2/3$. 
In this log-log plot, however, the exponent appears a bit smaller than $2/3$, approximately $0.6$. 
This may again be due to the off-set before reaching the 
scaling regime. In the inset of this figure we have shown $\zeta_i$ as a function of $1/m$, for $d=1$. 
The  convergence appears closer to
 $2/3$. For the sake of clarity, we avoided showing similar results from $d=2$ and $3$, which show similar trend in the direct plot
(at late time). We mention here that because of strong finite-size effects \cite{paul2_5} and difficulty 
in dealing with very large systems, the scaling regime is relatively small for $d=3$.
\par
~~This weak dependence of growth of mass on dimension not only invalidates equivalence between GGM and BAM in $d>1$, 
it also suggests the absence of any hyperscaling relation of the type obeyed by the BAM results. 
We mention here that choices of different $e$ or overall density do not alter the outcome. This fact is demonstrated 
in Fig. \ref{fig12}(b) for the $d=2$ case. There we have presented results for different $e$ and $\phi$ values. These 
results are consistent with the finite-size scaling estimate of exponent ($\simeq 0.3$) for the growth of 
average domain length \cite{paul2_5}.

\section{ Conclusion}
~~Via event-driven molecular dynamics simulations we have studied nonequilibrium dynamics in ballistic aggregation (BAM) \cite{carne_5, trizac1_5}
and granular gas (GGM) \cite{gold_5} models. We have presented accurate results on the energy decay and the growth of mass. These results are 
compared with the available theoretical predictions \cite{carne_5, trizac1_5}.
\par
~~We observe that for both the models the above mentioned quantities exhibit power-law behavior as a function of time. For 
the BAM, the corresponding exponents exhibit density dependence. Nevertheless, these exponents satisfy a 
hyperscaling relation \cite{trizac1_5}. With the increase of density, the energy and mass get inversely related to each other, the exponent 
being strongly dimension dependent. This latter observation is consistent with the prediction of CPY \cite{carne_5}. As a physical reason behind 
the difference between the exponents for energy decay and cluster growth in low packing fraction scenario, Trizac and Krapivsky \cite{trizac2_5} 
showed that, in this 
limit, the particles with kinetic energy larger than the mean undergo very frequent collisions, that enhances the dissipation.
\par
~~For the GGM we observe that the energy decay satisfies the prediction of CPY at all dimensions \cite{carne_5}. 
However, this is not always inversely related with the growth of mass. In fact the latter exhibits very weak dimension dependence. This 
aspect requires adequate attention. Furthermore, the exponent in this case does not match any known value for coarsening related to conserved 
order parameter dynamics \cite{bray_5}, with \cite{sigg_5} or without \cite{lifsh_5} hydrodynamics \cite{hansen_5}.
\par
~~In the context of BAM, further interesting studies are related to aggregation and fragmentation \cite{bril2_5, matt_5, rai_5}. 
Incorporation of fragmentation indeed can 
provide information on more realistic scenario. In future we intend to undertake comprehensive studies by including this fact.

~~{\bf Acknowledgment:} The authors thank Department of Science and Technology, Government of India, for financial support.
SKD also acknowledges the Marie Curie Actions Plan of European Commission 
(FP7-PEOPLE-2013-IRSES grant No. 612707, DIONICOS) as well as International Centre for
Theoretical Physics, Trieste, for partial supports.
SP is thankful to UGC, Government of India, for research fellowship.

~${*}$ das@jncasr.ac.in

\end{document}